\documentclass[]{spie}  
\usepackage[]{graphicx}
\usepackage{aas_macros}
    
\newcommand{\etal}{{\it et~al.\/}}

\def\comma{\, ,}

\def\vec#1{{\bf #1}}

\def\t0{\mbox{$t_0$}}
\def\r0{\mbox{$r_0$}}

\newcommand\insertfigFH[3][1.0]{%
  \begin{figure}
    \centering
      \includegraphics[height=#1\textheight,clip=true]{figs/#2}
    \caption{#3}
    \label{fig:#2}
  \end{figure}}

\title{Fringe tracking and spatial filtering: phase jumps and dropouts}
\author{David F. Buscher,
  John S. Young,
  Fabien Baron, 
  Christopher A. Haniff 
\skiplinehalf
Cavendish Laboratory, J. J. Thomson Avenue, Cambridge CB3 0HE, U.K.
}
\authorinfo{D.F.B.: E-mail: dfb@mrao.cam.ac.uk, Telephone: +44 (0)1223 337302}

\begin{document}
\def\preface{
Copyright 2008 Society of Photo-Optical Instrumentation Engineers.\\

This paper was (will be) published in Proc. SPIE 7013 and is made available
as an electronic reprint (preprint) with permission of SPIE. One print or
electronic copy may be made for personal use only. Systematic or multiple
reproduction, distribution to multiple locations via electronic or other means,
duplication of any material in this paper for a fee or for commercial purposes,
or modification of the content of the paper are prohibited.

\pagebreak
}
\pagestyle{plain}    

  \maketitle 

\begin{abstract}
  Fringe tracking in interferometers is typically analyzed with the
  implicit assumption that there is a single phase associated with
  each telescope in the array. If the telescopes have apertures
  significantly larger than $r_0$ and only partial adaptive optics
  correction, then the phase measured by a fringe sensor may differ
  significantly from the ``piston'' component of the aperture phase. In
  some cases, speckle noise will cause ``branch points'' in the measured
  phase as a function of time, causing large and sudden jumps in the
  phase. We present simulations showing these effects in order to
  understand their implications for the design of fringe tracking algorithms.
\end{abstract}
\section{Introduction}
A central component of any separated-element interferometer is the
system which tracks and corrects the time-varying phase differences between different apertures in
the interferometer. Typically, these systems are called ``fringe
trackers'' because they measure the phase of an interference fringe pattern, and
they measure and
correct only one spatial degree of freedom per aperture. This degree
of freedom is
commonly assumed to be simply the ``piston'' mode of the wavefront
error across the aperture, corresponding to the average wavefront
error, and the correction is applied using an
actuator capable of introducing piston, for example a
delay line.
 
Fringe-tracking systems have been operated
successfully in interferometers such as the Mark~III
interferometer\cite{Shao1980} and PTI\cite{Colavita1999}, and indeed
were critical to their scientific success. Both these interferometers had apertures which
were small compared with $r_0$ at the operating wavelength, and this means
that the approximation that there was only one atmospheric spatial
phase perturbation associated with each aperture was justifiable. The
latest generation of arrays such as CHARA\cite{Brummelaar2005}, Keck
Interferometer\cite{Colavita2004} and VLTI\cite{Glindemann2003}, all
have apertures which are much larger than $r_0$, and so it is
important to consider what the effects might be of
having significant ($>$1radian) perturbations across the apertures of each
telescope. 

Earlier investigations by Tubbs \cite{Tubbs2004,Tubbs2005}
showed that, in multi-$r_0$ interferometers with
beam combiners incorporating spatial filters, the interaction between the
focal plane ``speckles'' and the spatial filter
gave rise to substantial differences between the piston component
of the wavefronts and the phase of the wavefront transmitted through a spatial
filter. Studies by G. Daigne (private communication) showed that the
combination of spatial wavefront errors and diffraction effects in the
interferometric beam train could give rise to substantial phase
anomalies between the piston phase and the phase as measured in an
interference fringe. 

Here we tackle the question as to how and why spatial
wavefront corrugations can give rise to differences between the piston
phase and the phase measured by a beam combiner. We show that these
effects are quite general to all multi-$r_0$ interferometers 
and can arise whether or not diffraction or
spatial filtering are present in the instrumental setup. We
investigate briefly how spatial filtering and adaptive optics affect
the character of these phase anomalies, in order to understand how
best to design interferometers and their associated fringe-tracking systems.

\section{Measurement model}
We adopt here a simple model of an interferometer which is
nevertheless capable of
exhibiting the effects of spatial perturbations. The interferometer
consists of two telescopes with 
circular apertures of diameter $D$ observing a
distant unresolved 
source. The stellar wavefronts pass through an atmospheric phase screen containing
Kolmogorov-Tatarski spatial phase perturbations with Fried parameter
$r_0$, and the
phase screen is assumed to evolve with time as predicted by the Taylor hypothesis, with a single
phase screen moving at velocity $\vec v$ across the apertures --- more complex
models with multiple layers moving in different directions were not
considered here. The apertures are modeled
as being sufficiently far apart that the phase screens over each aperture
are uncorrelated, but that the wind velocity and direction are the
same for both apertures. 

The telescopes each have an adaptive optics
(AO) system. The AO is modeled as a perfect modal filter, which is to
say that the
first $n$ Zernike modes (excluding piston) are removed from the
wavefront. Values of $n$ from 2 to 20 are considered here,
where case $n=2$ corresponds to a tip/tilt correction system while
$n=20$ corresponds to a fifth-radial-order system.

The wavefronts from the telescopes are modeled as propagating without
diffraction or further degradation to a fringe-tracking sensor. For the purposes of this paper, the sensor is modeled as a
beam combiner which makes a quasi-monochromatic fringe pattern and
uses the phase of the detected fringes as a measure of the piston
difference between the two telescopes. Phase changes of more than one
wavelength are assumed to be tracked using ``phase unwrapping''
techniques, i.e. following the phase evolution with time and resolving
any $2\pi$ ambiguities in the phase by assuming that the phase changes
are continuous with time. We make no attempt here to model
beam combiners with multiple spectral channels or using envelope or
group-delay tracking methods, but we consider the implications of
such methods in the discussion.

Two sorts of beam combiner are considered. For an ``unfiltered''
combiner, the wavefronts from the two telescopes are overlapped in
either the pupil or the image plane, and fringes are detected either
through spatial and/or temporal modulation of the fringes. All the
light collected by the telescopes is detected: no spatial
filtering takes place. 

In a ``spatially filtered'' combiner, the
wavefront is subject to spatial filtering either before or after beam
combination (we assume that the spatial filter is identical for both
beams, so whether filtering occurs before or after combination makes
no difference to the resulting signal). Two forms of spatial filtering
are modeled: in ``pinhole'' spatial filtering we assume that a
pinhole is placed in a focal plane somewhere in the optical system, so that only
light from the central speckle can pass through. The
diameter of the pinhole is assumed to be much less than $\lambda/D$
where $\lambda$ is the wavelength of the radiation being observed (in
practice a pinhole diameter which is approximately $\lambda/D$
is to be preferred at low light levels\cite{Keen2001}, but the
approximation of an infinitely small pinhole used here 
simplifies the mathematics considerably). 

In ``fiber'' spatial
filtering, we assume a single-mode fiber or other mono-mode
waveguide is used to spatially filter the wavefront. The mode accepted
by the filter is assumed to be Gaussian-shaped and the f-ratio of the
optical system
coupling the light from the aperture into the waveguide is assumed to
be such that the diameter of the contour of the
far-field mode
profile where the mode amplitude drops to $1/e$ of the peak amplitude is 0.9 times
the aperture diameter, corresponding to optimal coupling
to the aperture\cite{Shaklan1987}.

Fringe detection is assumed to be noiseless, i.e. the source is bright
so there is no detector
or photon noise, so that any errors in the measured phase arise purely
from atmospheric distortions. 

\section{Simplified analysis}
In this section we attempt to gain some insight into what is measured
by an idealized fringe sensor and why and under what circumstances
this differs from the piston component of the wavefront. We consider
only the unfiltered combiner and the pinhole-filtered combiner, since
analysis of the fiber-filtered combiner is more complex and does not
yield any qualitatively new insights.

If we consider the fringes formed by an unfiltered beam combiner
(whether pupil-plane or image-plane), it is straightforward to
show\cite{Buscher1988} that the measured fringe pattern can be
characterized by a complex fringe amplitude given by
\begin{equation}
A=\int_S e^{i[\phi_1(\vec x)-\phi_2(\vec x)]}\,dS
\label{eq:unfiltered}
\end{equation}
where $\phi_1(\vec x)$ and $\phi_2(\vec x)$ are the phase
perturbations at position $\vec x$ within apertures 1 and 2
respectively (we define the
coordinate systems separately for the two apertures such that the value of
$\vec x$ represents the position with respect to the center of
the relevant aperture), $dS$ is an elemental within the aperture and the
integral over $S$
denotes integrating over the clear area of the aperture.
It is similarly straightforward to show that, for a
filtered combiner, the measured fringe complex amplitude is given by
\begin{equation}
A=A_1A_2^*
\end{equation}
where $A_1$ is the complex amplitude of the beam from aperture
1 after it has been spatially filtered. In the case of an infinitely
small pinhole spatial filter $A_1$
is given by:
\begin{equation}
A_1=\int_Se^{i\phi_1(\vec x)}\,dS\comma
\label{eq:pinholesingle}
\end{equation}
so that
\begin{equation}
A=\int_Se^{i\phi_1(\vec x)}\,dS\int_Se^{-i\phi_2(\vec x)}\,dS
\label{eq:pinhole}
\end{equation}

In both the filtered and unfiltered cases the
measured phase will be the argument of the measured complex fringe
amplitude $A$. We can compare this with the expression for the piston
component of the wavefront across each of the apertures. The piston
component of the phase across aperture 1 will be given by:
\begin{equation}
\Phi_1=\int_S\phi_1(\vec x)\,dS
\label{eq:pistonsingle}
\end{equation}
and the piston difference between the apertures will be given by
\begin{equation}
\Phi=\Phi_1-\Phi_2=\int_S\phi_1(\vec x)-\phi_2(\vec x)\,dS
\label{eq:piston}
\end{equation}

It is readily apparent that, while there is some similarity
between, for example, equation~\ref{eq:piston} and the phase of the
complex amplitude given in equation~\ref{eq:unfiltered}, it is by
no means obvious that they should in general give the same answer.

However, it is clear that they do measure the same quantity
in the limit of no phase fluctuation across the aperture, and it is
also possible to derive the relationship between the two for the case
when there are only small amounts of phase perturbations. If the phase is approximately constant across the aperture such that
\begin{eqnarray}
\phi_1(x)=\phi_1+\delta\phi_1(\vec x)\\
\phi_1(x)=\phi_2+\delta\phi_2(\vec x)
\end{eqnarray}
where $\delta\phi_1(\vec x)<<1$ and $\delta\phi_2(\vec x)<<1$, then we
can do a Taylor expansion of the complex exponentials in the above expressions. For example, we
can derive from equation~\ref{eq:pinholesingle} that
\begin{equation}
A_1\approx e^{i\phi_1}\left[1-\frac{1}{2}\int_S\left\{\delta\phi_1(\vec x)\right\}^2\,dS+i\int_S\delta\phi_1(\vec x)\,dS-\frac{i}{6}\int_S\left\{\delta\phi_1(\vec x)\right\}^3\,dS\right]
\label{eq:pinholecomplex}
\end{equation}
and by considering the real and imaginary components of the term in
the square brackets, we can see that the phase of this quantity is given approximately by
\begin{equation}
{\rm arg}(A_1)\approx\phi_1+{
\int_S\delta\phi_1(\vec
x)\,dS-\frac{1}{6}\int_S\left\{\delta\phi_1(\vec
  x)\right\}^3\,dS
\over
1-\frac{1}{2}\int_S\left\{\delta\phi_1(\vec x)\right\}^2\,dS}
\label{eq:pinholephase}
\end{equation}
We can see that, to first order in
$\delta\phi_1$, the phase of $A_1$ is indeed equal to the piston
phase. Furthermore, the deviation between the piston phase and the
argument of the complex amplitude will scale approximately as $\delta\phi^3$. 

We can
derive similar results for the phases of the filtered and unfiltered
fringes, namely that the fringe phase accurately represents the piston
phase when the instantaneous phase fluctuations across the aperture are small when
compared to a radian, but that as the phase fluctuations approach a
radian, there will be a rapid divergence between the piston phase and
the measured fringe phase.

When the aperture phase deviations $\delta\phi$ become large compared to a
radian, it is possible for large divergences between the fringe phase
and the piston phase to appear. What is more, these divergences can
give rise to
behavior which can be termed chaotic, by which we mean that relatively small changes
in the distribution of phase values in the aperture can give rise to
large differences in the final outcome. 

\begin{figure}
\begin{tabular}{ccc}
\includegraphics[width=0.3\textwidth]{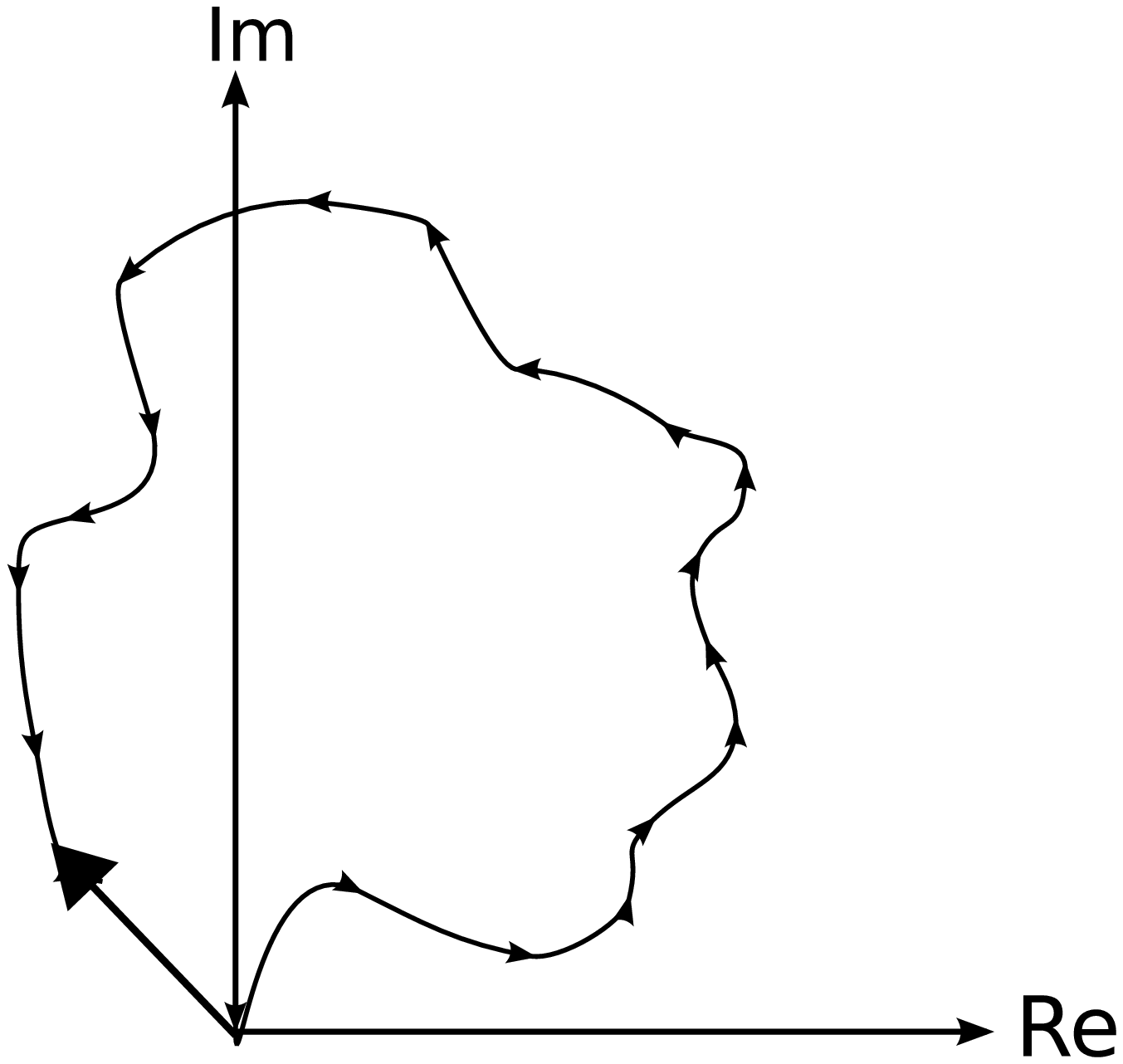} &
\includegraphics[width=0.3\textwidth]{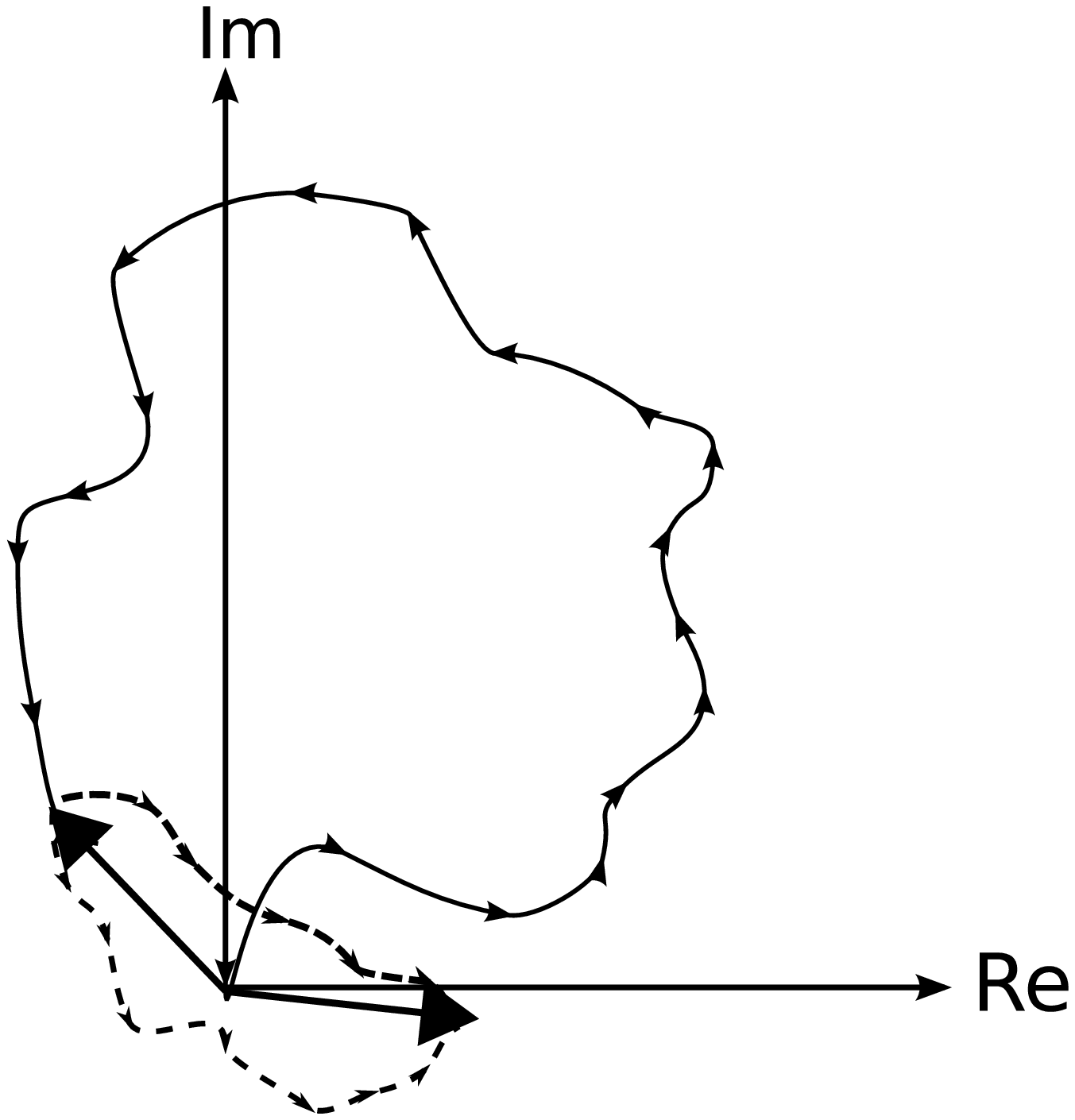} &
\includegraphics[width=0.3\textwidth]{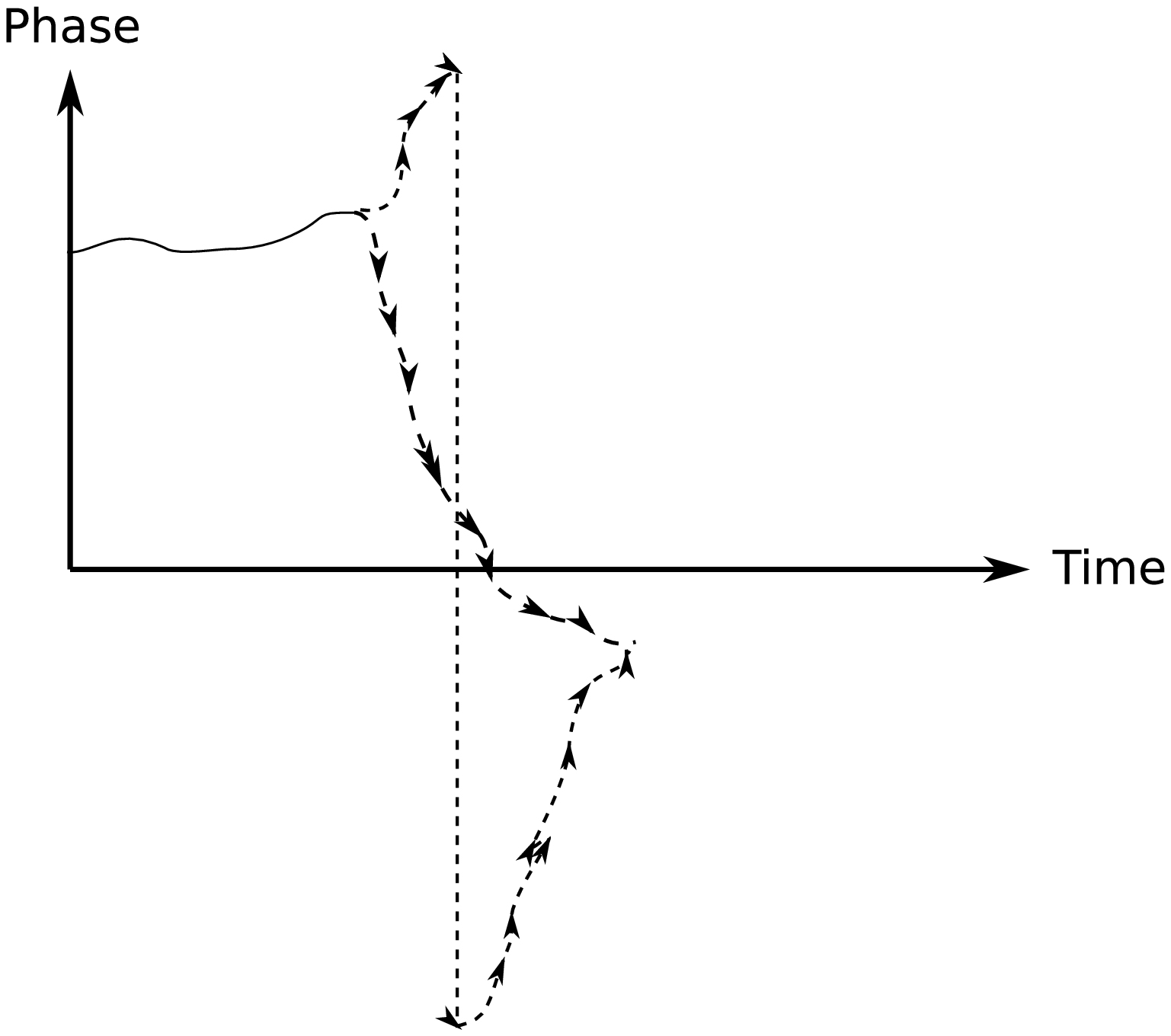} \\
(a) & (b) & (c)
\end{tabular}
\caption{(a) A complex amplitude summation in the complex plane,
  showing a low amplitude resultant phasor (b) Two possible subsequent
  evolution paths (shown as dotted lines) ending up at the same
  resultant phasor (c) The corresponding phase evolution showing a
  phase ``branch point''. The principal value of the phase is plotted,
  i.e. the phase jumps between $+\pi$ and $-\pi$ to keep the value
  within range.  }
\label{fig:branch}
\end{figure}

To see this, let us consider
the wavefront across an aperture at two instants in time, which are
separated by an interval which is small compared to $D/v$, the time taken for
the wind to sweep the phase screen across the aperture.
At any one instant
we can treat
the integral across the aperture in equation~\ref{eq:pinholesingle} as the limit of a sum
of phasors in the complex plane. If the range of
phase values is large compared to a radian, we can end up with the
resulting sum being of small amplitude, i.e. close to the origin of
the complex plane as
shown in Figure~\ref{fig:branch}(a). 

The low modulus corresponds to an instant of
low Strehl ratio in the focal plane image, but at the same time
corresponds to a regime of extreme sensitivity of the phase to small
perturbations. For example, at a time $\Delta t << D/v$ later, the distribution of
phases across the aperture will have changed by only a small amount:
if the Taylor hypothesis holds, approximately $1-\Delta t v/D$ of phasors
will be identical and only the fraction $v\Delta t/D$ of the phasors
will be different. If we track this small change in time, then two
possible evolutionary paths are shown Figure~\ref{fig:branch}(b). It
can be seen that two small and nearly identical evolutionary histories
for the aperture phases give two very different tracks in the fringe
phase as shown in Figure~\ref{fig:branch}(c), despite the fact that the piston phase ends up at the same
point.

We can see from this example that it is possible under conditions of
large phase perturbations to generate rapid ``jumps'' in the fringe
phase which are not seen in the piston phase. Furthermore, these jumps
can cause a $2\pi$ ``wrap around'' in the fringe phase which will
confuse any ``phase unwrapping'' algorithm in a similar to the phase
``branch points'' seen in 2-D phase maps.

In summary then, we have shown that the piston phase and fringe phase
will track each other closely if the instantaneous phase perturbations
across the aperture are small compared with a radian, but that large
differences can appear as the fluctuations approach and exceed a
radian.
 
\section{Simulations}
\subsection{Simulation method}
In order to demonstrate these effects numerically, a computer simulation of the
interferometer model was set up. Kolmogorov-Tatarski phase screens were generated on a
regular grid by
Fourier-transforming filtered white noise\cite{Buscher1988}. The grid
sampled the phase at intervals of $r_0/6$ or less and the grid extent was
typically greater than $160r_0$. Time series of Taylor screens were
generated by ``sliding'' the telescope aperture across the phase
screen, interpolating the phase screen onto a
regular grid in the aperture plane. It was possible to
generate long non-repeating sequences of aperture phase snapshots by setting the
``wind'' velocity vector at 30$^\circ$ to the
grid axes and ``wrapping'' round when the aperture reached the edge of
the phase screen (this makes use of the cyclic edge-to-edge
continuity of the Fourier-generated phase screens). Statistically independent phase screens were used for each of the apertures.

Adaptive optics was simulated by projecting out Zernike modes from the
aperture phases and subtracting these modes.  Three different beam
combiners were simulated: an unfiltered beam combiner, a pinhole
combiner (with an infinitely small pinhole) and a fiber combiner. Time
series of the complex fringe amplitudes measured by these combiners
were generated, together with, in the case of the filtered combiners,
the complex amplitudes of each of the filtered beams.
The piston phase for
each aperture was calculated as well as a modified version of the
piston phase with a Gaussian weighting across the aperture. The latter
was used as a comparison with the fiber-filtered fringe data, so that effects
due to the aperture weighting of the fiber could be eliminated,
i.e. at small $D/r_0$ this modified piston phase and the fiber phase
should converge rapidly to one another.

\subsection{Phase jumps}
Figure~\ref{fig:phasesequence} shows the output of one of the
simulations. It shows a system with $D/\r0=4$ and tip/tilt correction,
where the mean Strehl ratio is greater than 0.3. Nevertheless, the
Strehl ratio shows frequent ``dropouts'' to much lower values; this is due only to
the statistical nature of the phase fluctuations as a perfect tip/tilt
system has been modeled. 

\insertfigFH[0.5]{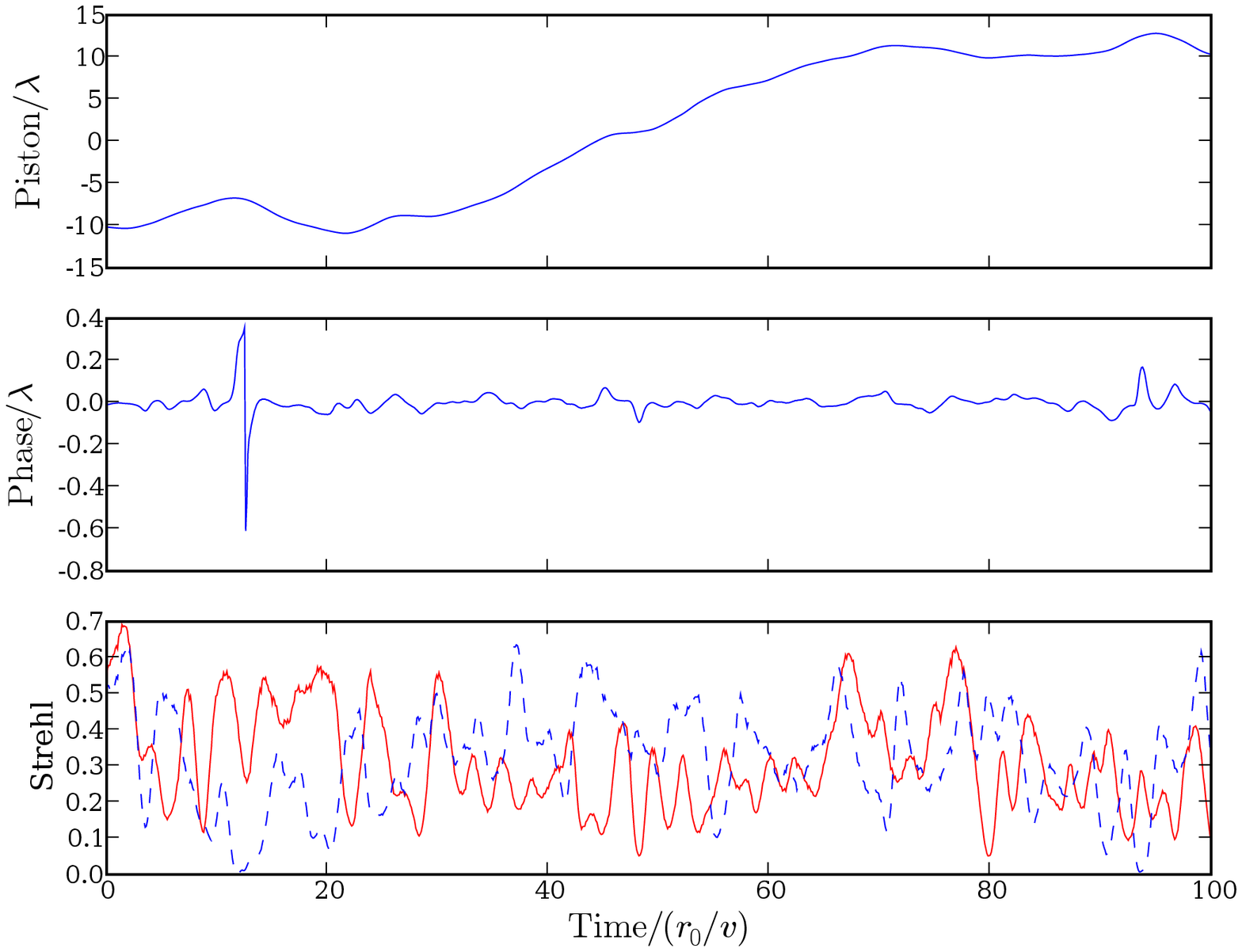}{Simulated time-sequences for an
  interferometer with tip/tilt correction and $D/\r0=4$. The top
  graph shows the measured piston phase, which shows the smoothing
  expected from ``aperture
  filtering''. Below it is the difference between the piston phase and
  the phase measured with a fiber-filtered beam combiner. The bottom
  graph shows the Strehl ratios seen at each of the two
  telescopes, shown as solid and dashed lines respectively. It can be seen that there are occasional large phase
  ``jumps'' associated with periods of low Strehl.}

For most of the time, the fringe phase tracks the piston phase to much
better than a radian, but on some (but not all) of the occasions where
the Strehl in one aperture or another falls to a low value, the phase
measured by the beam combiner shows large ``jumps'' where the measured
phase differs from the piston phase by more than a radian. These jumps
are fast: the time axis is plotted in units of $r_0/v$, and so for
$D/\r0=4$ we see that, as expected from ``aperture filtering'', the
piston phase shows no structure on timescales of less than 4 units,
while the phase jumps occur on timescales of 1-2 units.

This observation follows what is expected from the previous analysis:
when the Strehl is low, the measured phase can change rapidly and by
large amounts. Note that the Strehl being low does not inevitably lead
to large phase jumps, since the combination of low Strehl and a
suitable subsequent evolution of the aperture phase distribution are needed to
cause large phase fluctuations.

The combination of low Strehl and fast fluctuations represents a
double challenge for a fringe tracker, because the low Strehl ratio
implies that the signal-to-noise ratio on a fringe measurement is
reduced. At the same time, the phase is changing rapidly, so the
tracker must sample fast to keep up. 
Even if the tracker is able to meet these stringent requirements, we
can see that it will still perform less than adequately on some
occasions: in the first large jump in phase seen in
Figure~\ref{fig:phasesequence}, we can see that the phase difference
between the fringe-tracker phase and the piston phase wraps through
$2\pi$ radians (i.e.\ we have gone through a ``branch point''), so the
fringe tracker will end up one fringe off the fringe it was initially
following. If these jumps continue, the tracker could eventually end
up losing the fringe packet altogether.

The frequency of such large jumps depends on the $D/\r0$ value for the
aperture. At $D/\r0=4$ such jumps are seen in almost every sequence of
the length shown in Figure~\ref{fig:phasesequence}: for typical
values at visible wavelengths of $r_0=10$cm and $v=10$m/s this
correspond to timescales of order 1 second. For lower $D/\r0$ values,
the rate of such jumps fall dramatically, such that at $D/\r0=3$ much
longer sequences are required to see the jumps. A statistical analysis
of the rate of these jumps as a function of aperture size will be the
subject of a later paper.   

It might be argued that the fact that the fringe tracker does not
follow the ``piston phase'' is irrelevant, because what is important
that we follow the phase seen by the science camera. However, if the
science camera is at a different wavelength to the fringe tracker,
then due to the non-linear nature of the phase jump phenomenon, the
phase jumps seen in the science camera and the fringe tracker are only
weakly correlated. Figure~\ref{fig:wavelengthcomparison} shows a
simulation where the wavelength ratio between the science camera and
fringe tracker is only 1.5 (it does not matter which way round the
ratio is for the purposes of this analysis), and yet the phase jumps
seen at one wavelength are rarely mimicked at the other wavelength.

\insertfigFH[0.3]{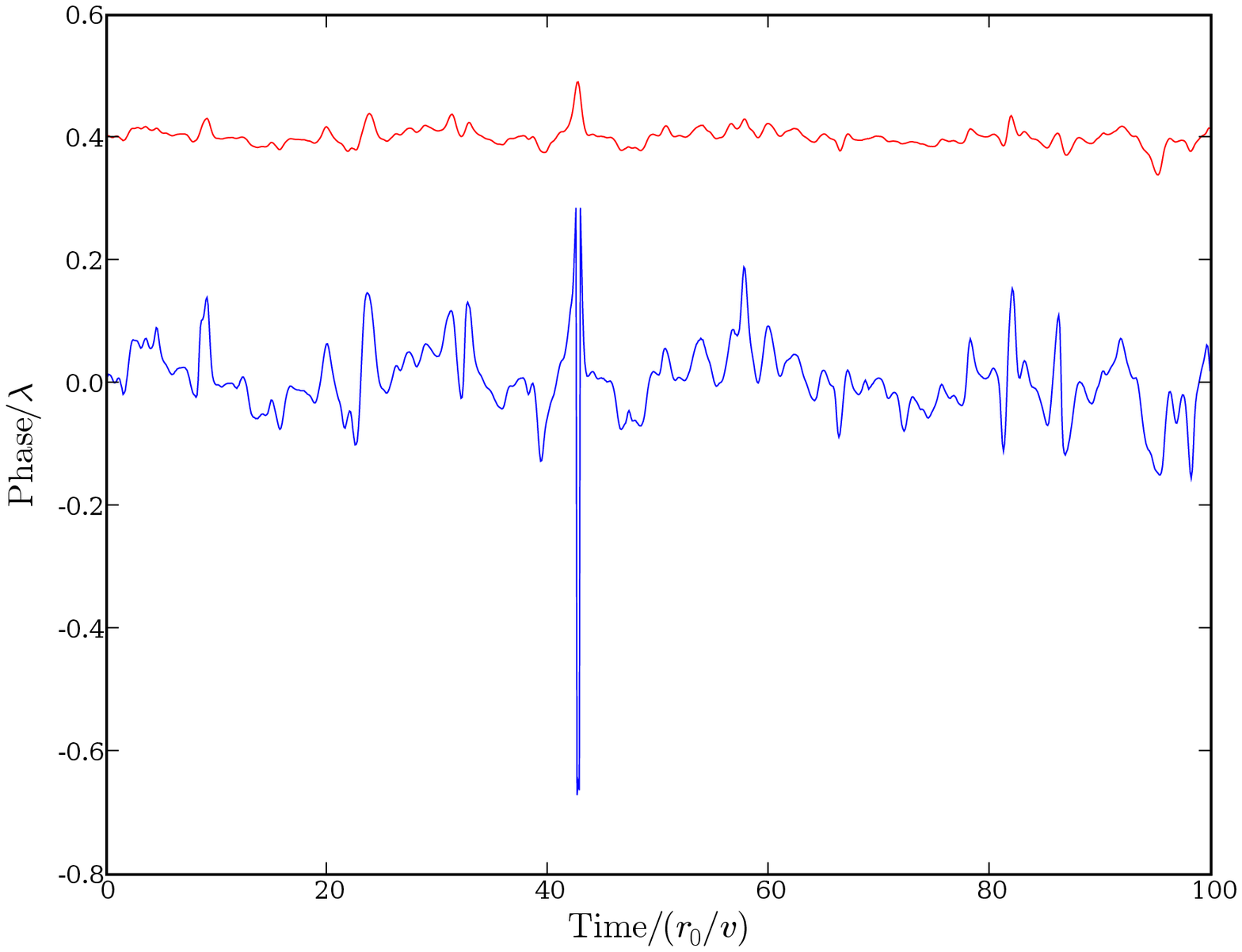}{Simulated phase sequences seen
in two beam combiners operating at different wavelengths. The upper
sequence (offset for clarity) is for a combiner operating at a $D/\r0$
value of 3, while the lower sequence is for a wavelength a factor of
1.5 times smaller. We can see that the magnitude of the phase jumps at
different wavelengths rarely if ever exhibit the ideal ratio of 1.5
and hence fringe tracking at one wavelength will not perfectly track
the phase jumps at another wavelength. The system has tip/tilt
correction only.}

\subsection{Spatial filtering}
An interesting question to ask is whether or not spatial filtering
improves or makes worse the differences between the piston phase and
the measured phase. The implication from Tubbs' work\cite{Tubbs2005}
was that spatial filtering might in fact be the source of the
anomalies, but we have already shown that this is not the case.

As we have seen, the phase anomalies are associated with non-linear effects which occur
when the the variation of the phase of the complex exponentials in
equations~\ref{eq:unfiltered} and \ref{eq:pinhole} become
comparable to a radian. It can be readily shown that the instantaneous 
variation of
the imaginary argument
to the exponential for the unfiltered combiner
(equation~\ref{eq:unfiltered}) will typically be larger by a factor of
order $\sqrt{2}$ than that for
the for the filtered combiner (equation~\ref{eq:pinhole}), and
therefore we expect the unfiltered combiner to be \emph{more} likely
to be affected
by phase anomalies than the filtered combiner: phase jumps correspond to the infrequent realizations in
the ``tail'' of the set of possible instantaneous
aperture phase distributions, and the chance of the instantaneous
fluctuations approaching or exceeding a radian is significantly greater for the
unfiltered combiner. The fact that
there are two ``chances'' (one for each integral in
equation~\ref{eq:pinhole}) to have an unusual distribution in the case
of the filtered combiner does not make up for the greater probability
of a large-variance event in the unfiltered combiner. 

The simulation bears out this analysis:
Figure~\ref{fig:filteringcomparison} shows that the phase jumps are both
larger and more frequent in the unfiltered combiner than in the
filtered combiner.

\insertfigFH[0.3]{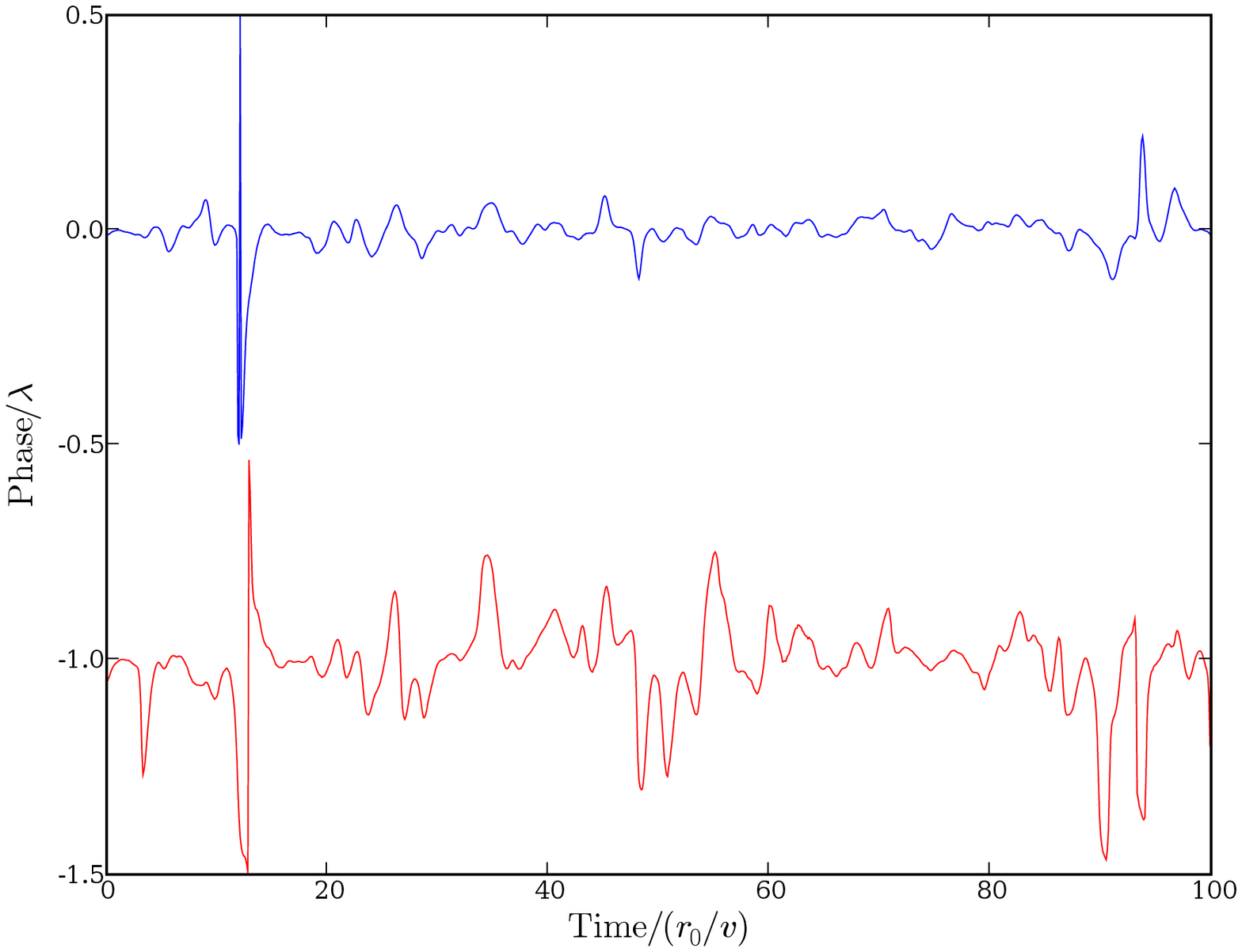}{A comparison of the phases
  measured by a pinhole-filtered beam combiner (top) and an unfiltered
  beam combiner (bottom), made using the same wavefronts. The
  simulated value of $D/\r0$ is 4, and the system is tip/tilt corrected.}

\subsection{The effect of adaptive optics}
The analysis presented earlier indicates that it is the size of the
spatial fluctuations that determines the strength of the anomalies
seen between the phase
measured by a fringe sensor and the piston phase. A higher-order AO
system operating on a given aperture size will therefore reduce the
likelihood of seeing phase anomalies. However, AO systems are
typically sized to the aperture they will be used on, so a more
interesting question is whether a large aperture system with
high-order AO is better than a small-aperture system with lower-order
AO.

To first order, we expect that the AO system will be designed to give a
certain minimum Strehl ratio for the aperture/science wavelength combination
it is being used with. Given that the Strehl is a monotonic function
of the RMS residual phase perturbations in the aperture, we might
expect apertures with different orders of AO but the same Strehl to
show similar phase-tracking anomalies. In fact, the simulations 
do not bear this out: the mean Strehl ratio at which 
higher-order AO systems first exhibit frequent phase
anomalies appears to be lower than for their lower-order
counterparts. 

One possible explanation lies in the spatial structure of the residual
phases: seen in the image plane, a higher-order system of the same
Strehl ratio will have a higher contrast between the central speckle
and the speckle ``halo''. Examination of simulated image data at the
moments when the
large phase jumps are seen shows that these occur when the
central speckle disappears and gives its energy to a new speckle. The
tip/tilt component of the AO system will tend to recenter on the new
speckle, making it the central speckle, which, at least in a
spatially-filtered system, then dominates the phase of the fringes. This new speckle will have a
completely different phase from the original central speckle and therefore give rise to a rapid jump in
the measured phase. The theory as to why higher-order systems are
better would then have to invoke the fact that such a ``speckle
exchange'' event is far more unlikely in the higher-order case,
because the chance of one of the subsidiary speckles having comparable
energy to the main speckle is finite.
This needs to be investigated further to determine if this theory
provides a consistent explanation for all the phenomena observed.

\section{Discussion and conclusions}
We have seen that discrepancies can arise 
between the phase measured by a fringe
sensor and the piston phase, and that these are a strong function of
the level of phase corrugations across the aperture. These
discrepancies are often seen in the form of large and rapid phase
``jumps'' and they occur preferentially when the instantaneous Strehl ratio or
the fringe amplitude is small, making these jumps hard for the fringe
tracker to follow. Indeed, if the fringe tracker does follow these
excursions, it may be doing the wrong thing, either
because it will ``phase wrap'' onto the wrong fringe, and/or will
follow a phase excursion which is not seen at the science wavelength.

Both spatial filtering and adaptive optics help to reduce the level of
such anomalies, but do not eliminate them. The only way to completely
avoid these anomalies is to work in a regime where these jumps are
unlikely, by selecting appropriate values for $D/\r0$ and the spatial
order of the AO system. In practice this is quite difficult to do:
while it is possible to adjust the effective aperture of the system
appropriate to the prevailing value of $\r0$ using a simple diaphragm,
there are two problems with this approach. Firstly, for most AO
systems it is difficult to change the effective aperture size
dynamically and still maintain the same spatial order of correction,
because this involves rescaling the spatial resolution of both the
wavefront sensor and the wavefront 
corrector to the new aperture size. Secondly, it is quite difficult to
maintain an optimum aperture size because of rapid variations in the
seeing. Measurements by Baldwin \etal\cite{Baldwin2008} show that, at least
on one (quite good) site and at some times, the spatial scale of the seeing can
vary by factors of more than 2 on timescales of a few hundred milliseconds
as shown in Figure~\ref{fig:baldwin}.

\insertfigFH[0.3]{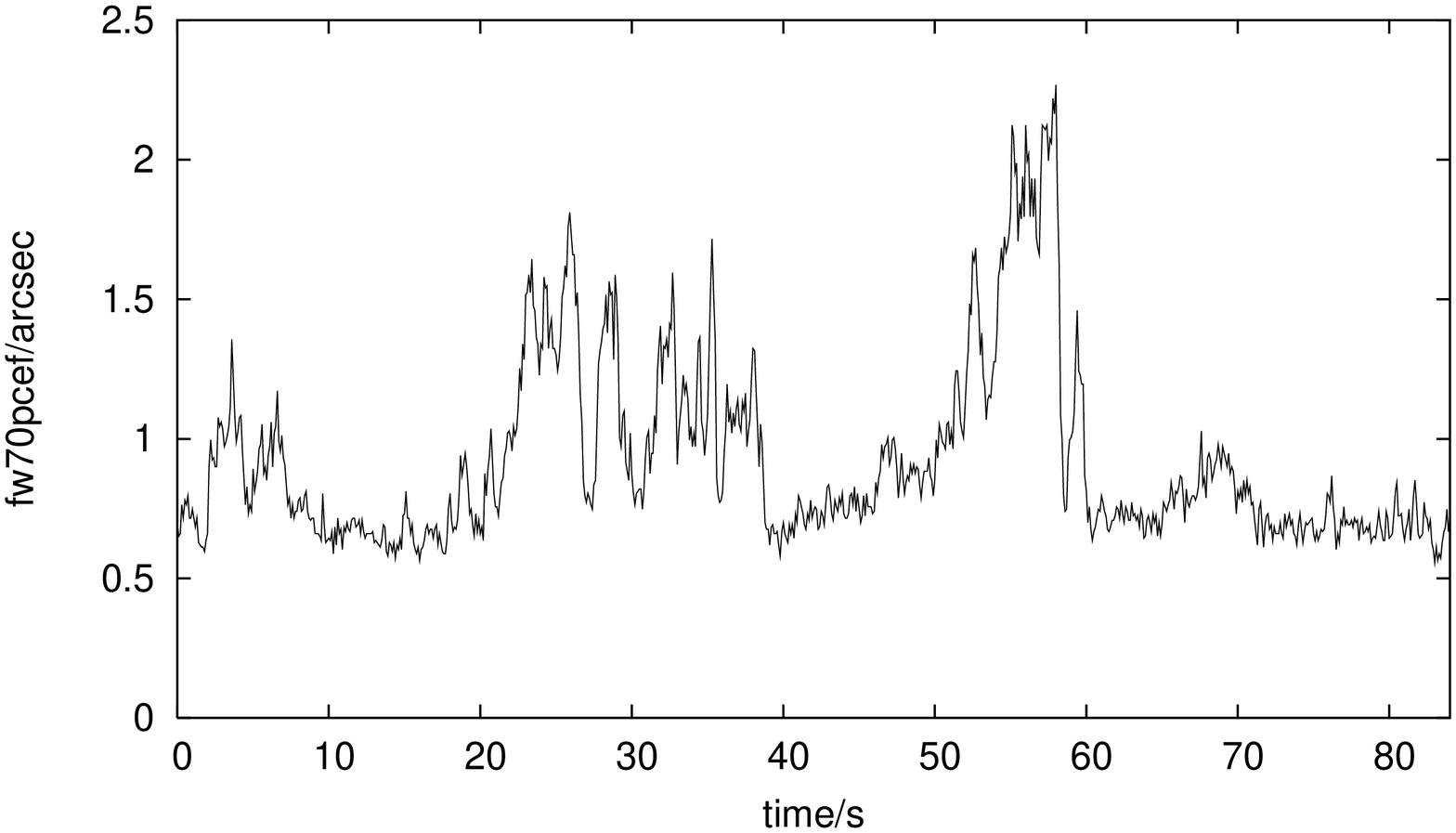}{Example of rapid variations in the seeing
  from Baldwin \etal\protect\cite{Baldwin2008} (reproduced with
  permission of the authors). The y-coordinate, {\tt fw70pcef}, is
  a measure of the intrinsic seeing in arcseconds which is accurate to about
  10\% at any instant. It can be seen that $\r0$ changes by a factor
  of more than 3 over very short timescales, with some changes
  occurring in less than a second.}

The above arguments suggest that all successful fringe trackers need to be able to deal
with rapid phase jumps. This requires development of robust
fringe-tracking algorithms
which either recognize the phase jumps when they occur (perhaps this may be possible
using auxiliary data from the AO system?), or recover rapidly
from the loss of fringe lock that may result. An alternative is to
use algorithms that are inherently insensitive to rapid phase
variations. An example is group-delay fringe tracking or 
any similar algorithm which tracks the coherence envelope of the
fringes rather than the fringe phase. These algorithms tend to average
the data over timescales of many tens of $\t0$, so that infrequent and
rapid
anomalies in the phase will have little effect. These algorithms have
the additional advantage that they are little affected by short
periods of poor seeing. However, the investigation here has only been
the effects on a phase-tracking system and not a group-delay system,
and so this is a fruitful area for future research.

\section{Acknowledgments} 
Our search for perfection in fringe tracking received considerable
input from H. Blumenthal, for which we acknowledge our deep
appreciation.


\bibliography{buscher-fringetracking}
\bibliographystyle{spiebib}   

\end{document}